\newcommand{\beq}{\begin{equation}}
\newcommand{\eeq}{\end{equation}}

\documentclass[aps,twocolumn,superscriptaddress,showpacs]{revtex4-1}
\usepackage{epsfig}
\usepackage{natbib}
\usepackage{amsmath}
\usepackage{times}
\usepackage{psfrag}
\usepackage{epstopdf}
\usepackage{braket}

\usepackage{graphicx}
\usepackage{subfigure}
\usepackage{float}
\begin{document}
\title{
Nonideal Quantum Measurement Effects on the Switching  Currents Distribution of Josephson Junctions
}
\author{Vincenzo Pierro}
\affiliation{Dept. of Engineering, University of Sannio, Corso Garibaldi, 107, I-82100 Benevento, Italy}
\author{Giovanni Filatrella }
\affiliation{Dept. of Sciences and Technologies, 
University of Sannio, Via Port'Arsa, 11, I-82100 Benevento, Italy}
\date{\today}

\begin{abstract}

The quantum character of Josephson junctions is ordinarily revealed through the analysis of the switching currents, i.e. the current at which a finite voltage appears:
A sharp rise of the voltage signals the passage (tunnel) from a trapped state (the zero voltage solution) to a running state (the finite voltage solution).
In this context, we investigate  the probability distribution of the Josephson junctions switching current taking into account the effect of the bias sweeping 
rate and introducing a simple nonideal quantum measurements scheme.
The measurements are modelled as repeated voltage samplings at discrete time intervals, that is with repeated projections of the time dependent quantum solutions on the static 
or the running states, to retrieve the probability distribution of the switching currents.
The distribution appears to be immune of the quantum Zeno effect, and it is close to, but distinguishable from, the Wentzel-Kramers-Brillouin approximation.
For energy barriers comparable to the quantum fundamental energy state and in the fast bias current ramp rate the difference is neat, and remains sizeable in the 
asymptotic slow rate limit. 
%We find that such effect is vanishingly small with the increasing of the energy barrier, disappearing in the limit where Josephson junction is well described by a cubic oscillator.
This behaviour is a consequence of the quantum character of the system that confirms the presence of a backreaction of quantum measurements on the outcome of mesoscopic 
Josephson junctions.
\end{abstract}
\maketitle

\section{Introduction}
\label{Introduction}
Josephson Junctions (JJ) are a well established playground for macroscopic quantum tunneling  \cite{Martinis87,Shnirman97,Martinis02,Wallraff03b,Price10,Coskun12,Massarotti15} and a prominent field where the basic concepts of quantum mechanics have been successfully demonstrated in a mesoscopic system, with potential technological implications in quantum computing \cite{Makhlin01}.
The subject is of interest per se as an application quantum dynamics to a macroscopic object (analogous, for example, to the particle-like dynamics of fluxons described by collective coordinates \cite{Shnirman97} or to the motion of the end masses of advanced gravitational wave detectors working near the quantum limit \cite{Abbott16}) and to  validate non ideal quantum measurements models \cite{Andersen13}.
On the mathematical side, the quantum dynamics associated to unbounded potentials, as the cubic approximation \cite{Caliceti80,Alvarez88}, is an open problem; possible approaches are the Wentzel-Kramers-Brillouin (WKB) approximation \cite{Martinis87,Tinkham96}, the semi classical approach \cite{Kivioja05} or the time dependent imaginary potential  \cite{Andersen13}.
On the experimental side, the quantum character of the dynamics of Josephson devices has been investigated in pioneering works that have shown the occurrence of tunnel, or the passage across a barrier higher than the available system energy \cite{Martinis87}. 
The quantum nature of the phase difference is in fact uncovered by showing that no matter how cold is the junction, the phase difference can overcame a finite barrier energy and move from the zero to the finite voltage state.
The experiments are delicate, and the mere appearance of a switch event (a sudden appearance of a finite voltage) at a temperature where it is unlikely, is not a definitive proof.
In fact one could conceive that the environment noise enters the system bearing an effective temperature that is higher than the thermodynamic temperature.
When escapes can be ascribed to stochastic activation, it is difficult to discern between thermal activation \cite{Buttiker83,Augello09} and quantum tunnel\cite{Martinis87,Martinis02,Wallraff03b,Price10,Coskun12,Massarotti15}, inasmuch several quantum effects, as resonance with level quantization \cite{Groenbeck04} and Rabi oscillations \cite{Groenbeck05}, have been reproduced in classical activated JJ.
It is therefore important to pinpoint effects that are unique for quantum systems, and could not possibly be confused with classical analogue.
In this work we propose to consider the effect of non ideal (non von Neumann) quantum measurements \cite{Barchielli09}, as the influence of measurements on the system is a peculiar feature of quantum mechanics, without a classical counterpart.
Repeated measurements on a quantum system (monitoring) require that a detector reveals the position, and that such determination affects the subsequent dynamics of the system, inasmuch the quantum track associated to the measurements process entails a sequence of evolution and projections on the appropriated subspace \cite{Anastopoulos08,Dhar15,Konrad10}.
In this context, we apply a discrete time scheme, where measurements only take place at discrete, predetermined (by the observer) time intervals.
We choose discrete time measurements also to avoid the problems connected with continuous time monitoring of a quantum system  \cite{Anastopoulos08,Dhar15,Wiseman93} and the quantum Zeno effect \cite{Abdullaev11,Abdo14}.
Discrete measurements allow to investigate whether this extra degree of freedom, the frequency of the measurements, has an impact \cite{Matta15} on quantities experimentally accessible for JJ, namely the switching currents.
Our investigation points toward a qualitative effect of the measurements frequency: the peak of the probability distribution of the exit currents,  moves to lower values.

The work is organized as follows. In Sect. \ref{Model} we set the stage for the analysis of quantum measurements, first describing in Subsect. \ref{equations} 
the basic equations and the appropriated boundary conditions to hinder spurious reflections, then  in Subsect. \ref{switching} we describe a procedure to project the two outcomes of the measurements (either zero or nonzero voltage) for the calculation of the probability distribution of the switching currents.
In Sect. \ref{Dynamics} we detail the effect of the measurements on the zero voltage solution.
Sect. \ref{Results} contains the main outcome of the approach: the effect of the measurement frequency and of the bias current ramp time on the probability 
distribution of the switching currents. After the Conclusions of Sect. \ref{Conclusions}, 
the Appendix details the dependence of the distribution upon the sweep bias time in the adiabatic, WKB approximation.

%%%%%%%%%%%%%%%%%%%%%%%%%%%%%%%%%%%%%%%%%%%%%%%
\section{The model}
\label{Model}
In this Section we describe the basic equations, included the treatment of boundary conditions, in Subsect.  \ref{equations}. 
The procedure to retrieve the probability distribution of the switching currents from the projection procedure is in Subsect. \ref{switching}.

%%%%%%%%%%%%%%%%%%%%%%%%%%%%%%%%%%%%%%
\subsection{Basic equations}
\label{equations}
The quantum dynamics governing the gauge-invariant phase difference $\varphi$ between two superconductors is given by a probability distribution $\psi (\varphi,t')$, 
as results from the solution of a Schr\"odinger equation: 
\beq
i\hbar \frac{\partial \psi}{\partial t'} = \left[ - \frac{\hbar ^2}{2M}\frac{\partial^2}{\partial \varphi^2} -  E_J \left(  \cos (\varphi ) +  
\gamma(t') \varphi \right)  \right] \psi.
\label{eq:JJSE}
\eeq
in normalized units \cite{Tinkham96,Andersen13} it reads:
\beq
i \frac{\partial \psi}{\partial t} = \left[ -\frac{1}{2}\frac{\partial^2}{\partial \varphi^2} -  V_0 \left( \cos\left(\varphi\right) +  \gamma (t) \varphi \right) \right] \psi.
\label{eq:JJSEn}
\eeq
\noindent Here time is normalized to $\hbar/M$, where for a small JJ the mass reads $M= C\left( \Phi_0 /2\pi \right)^2$, where $C$  is the junction  capacitance and $\Phi_0= h/2e$ 
denotes the magnetic flux quantum. 
The dimensionless parameter $V_0=E_J/E_C$ is the normalized maximum energy barrier that results from the ratio between the Josephson energy $E_J=I_0\Phi_0/2\pi$ and the
 Coulomb energy $E_C=\hbar^2/[C\left(\Phi_0/2\pi\right)]^2$. Where $I_0$ is
the critical current of the JJ. Normalizing the bias current with respect to $I_0$, the time dependent applied current $I (t)$ reads $\gamma(t)= I(t)/I_0$, 
that  is ramped in a time $T$ from $\gamma(0)=0$ to $\gamma(T)=1$. 
To increase the normalized bias current $\gamma$ amounts to tilt the potential 
\beq
U=-V_0 \left[  \cos(\varphi) + \gamma \varphi \right]
\label{eq:U}
\eeq
associated to Eq.(\ref{eq:JJSEn}), while the corresponding barrier (normalized respect to $E_C$)
\beq
\label{eq:potential}
\Delta U =2 V_0 \left[ \sqrt{1-\gamma}-\gamma  \arccos{\gamma}\right]
\eeq
decreases and eventually, close to $\gamma=1$, becomes small enough to make the tunnel probability current sizable. 
The process is initiated for $\gamma=0$  when the Hamiltonian (\ref{eq:JJSEn}) is periodic; at this initial point it is appropriated to use the eigenfunctions:
\beq
\label{eq:ICa}
\psi(\varphi , 0 ) = \frac{ 1 }{\sqrt{2\pi}} {\mathrm c e}_{0}[ (\varphi + \pi) /2, 4V_0  ],
\eeq
where $  {\mathrm c e_{2n}}$ are $c e$ Mathieu's cosine elliptic functions \cite{Abramowitz72,Leibschner09}.
For any subsequent time, even after a negligibly lag, the bias current breaks the symmetry and destroys the periodicity. 
Consequently, the wavefunction collapses into the truncated equation:
\beq
\label{eq:IC}
\psi(\varphi , 0^+ ) = \frac{ \theta\left( \varphi+\pi\right) - \theta \left(\varphi - \pi \right) }{\sqrt{2\pi}} {\mathrm c e}_{0}[ (\varphi + \pi) /2, 4V_0  ].
\eeq
Here $\theta$ is the Heaviside step function.
The boundary conditions for the zero current case are periodic, $\psi(-\pi, 0) = \psi(\pi , 0)$, while for any finite current  the appropriate boundary conditions
\beq
 \forall t>0:  \lim_{\varphi \rightarrow \pm \infty} \psi \left(\varphi , t \right ) = 0.
\label{eq:bc_zero}
\eeq
are such that the wavefunction vanishes at $\pm\infty$. 
The limit at $-\infty$ is physically ensured, for the increasing potential forbids propagation.
More care is necessary at the edge at $+ \infty$, for the potential unbounded from below requires an absorbing condition that determines a net probability flux at infinity.
%\beq
%\lim_{\varphi \rightarrow \infty} \left( \frac{\partial  \psi(\varphi)}{\partial \varphi} -ik \psi(\varphi) \right)  = 0,
%\label{eq:bc_absorbing}
%\eeq
%with an appropriated choice of $k$ 
In principle the integration extends indefinitely in space, but for numerical simulations, that we perform with a Cranck-Nicholson method \cite{Numrecip}, it is of course necessary to truncate the domain.
This truncation is problematic, inasmuch an abrupt discontinuity generates spurious reflections that are incompatible with a running state towards infinity.
We have therefore inserted a perfect matched layer \cite{Taflove95,Oskooi08} that acts as an absorber at a finite distance and avoids unphysical backwards reflections from the finite distance boundary.
We denote with $P\left( \varphi < \infty , t \right)$ the probability to locate the solution in any finite position of the integration domain, that is not conserved in the presence of a perfect matched layer boundary \cite{Pierro16}.
The missed probability corresponds to the to {\em absorbed norm}  radiated towards infinity.

\subsection{Switching current probability distribution}
\label{switching}

The wavefunction $\psi(\varphi,t)$  exhibits the special features of a quantum solution, for instance it can tunnel through an energy barrier \cite{Martinis87}, or stay at discrete metastable energy levels \cite{Martinis87}, or show Rabi oscillations \cite{Martinis02}, to name few examples among many \cite{Wallraff03b,Price10,Massarotti15}.
These quantum effects are macroscopic, for the phase $\varphi$ is linked to the macroscopic current $\gamma$ and voltage $v$ through the celebrated Josephson equations:
\begin{eqnarray}
\label{eq:JJcurrentn}
\gamma &=& \sin \varphi, \\
\label{eq:JJvoltagen}
v &=& \frac{d\varphi}{dt}.
\end{eqnarray}

\begin{figure}
\centerline{\includegraphics [scale=0.39]{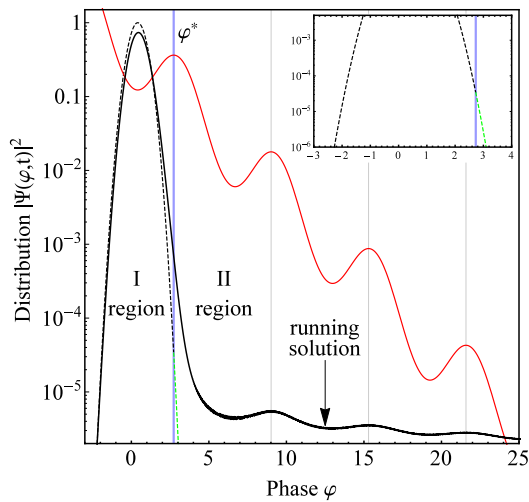}
%\hspace{1cm} 
%\includegraphics [scale=0.38]{Fig1b.eps}
}
%\label{fig:potential}
\caption{(color online) Sketch of the measurement process at the time $t=nT/N$.
The red solid curve sketches, in arbitrary units, the washboard potential, Eq.(\ref{eq:U}).
The vertical blue line corresponds to the separation $\varphi^*$ between the trapped state (region I) and the running state (region II).
The dashed black curve (also reported in the inset) except for a normalization constant is  the square modulus of the wavefunction, just after the measurement, $|\psi_2 (\varphi,t)|^2$, see Eq. (\ref{eq:IC}).
The solid curve represents its evolution, as per Eq.(\ref{eq:JJSEn}), after $10^3$ normalized units:   $|\psi_2 (\varphi,t+10^3)|^2$.
The presence of a finite probability in region II, behind the energy barrier limited by $\varphi^*$ corresponds to the possibility of {\em escapes}, i.e. to tunnel towards nonzero voltage solutions.
The non zero current of probability at the boundary demonstrates that the perfect matched layer \cite{Pierro16} gives the expected  decaying towards infinity. 
}
\label{fig:potential}
\end{figure}

It is the connection between the phase $\varphi$ and the accessible quantities $\gamma$ and $v$ that allows to detect tunnel events.
In fact, the experiments to ascertain the quantum nature of the JJ are based on the analysis of the distribution of the passages, or {\em switchings} (as often named in JJ jargon), to the finite voltage state \cite{Martinis87,Martinis02}, for the direct observation of the quantum phase difference is not possible. 
To illustrate the meaning of switching current, we refer to Fig. \ref{fig:potential}.
The phase described by the wavefunction $\psi_1(\varphi,t)$ just before the voltage measurement (denoted by a dashline in Fig. \ref{fig:potential}) is assumed to tunnel 
across the potential barrier if it is found in region $II$. 
If found in the region $\varphi > \varphi^*$ the representative point {\em runs downhill}: the phase suddenly increases and a voltage spike appears, as per Eq.(\ref{eq:JJvoltagen}). 
It is therefore possible to ascertain if a tunnel event (the passage from region $I$ to region $II$) has occurred measuring a voltage. 

%%%%%%%%%%%%___FIG___1___potential_____%%%%%%%
%%%%%%%%%%END FIGURE%%%%%%%%%%%%%%%%%%%%%%%%%%

To  include the effect of the measures at precise time intervals, we use the following simple model of non ideal quantum measurements.
Let us assume that in the ramp time $T$ one performs $N$ repeated measurements of the voltage at each instant $t=nT/N, \, n=1,2,...,N$.
When a measurement occurs, the two possible outcomes are: {\it i)} a finite voltage has been measured (switching), or {\it ii)} no voltage is detected (zero voltage state).
We identify the switching probability $p\left( t=nT/N  \right)$ of outcome {\it i)} with the probability to localize the wavefunction $\psi_1 (\varphi,t)$ behind the maximum of the potential, $\varphi > \varphi^*$:
\beq
\label{eq:p}
p \left( t=nT/N  \right) \equiv  \int_{\varphi^{*}}^{\infty}  | \psi_1 (\varphi,t) | ^2 d\varphi .
\eeq
Quantum mechanics dictates to model the effect of the measurement process on the wavefunction.
Following the standard von Neumann interpretation of measurements, one should project the state on the two corresponding subsets.
Unfortunately, the operator associated to the projections on the two measurements outcomes is not easily defined \cite{Andersen13} (the same difficulty that affects  arrival time quantum problems \cite{Dhar15}). 
To model the effect on the wavefunction $\psi_1 (\varphi, t)$ of a voltage detector that discriminates between static ($v=0$) and dynamics ($v\neq0$) we introduce a nonideal quantum measurement scheme as follows. 
If the JJ has moved from the supercurrent to the finite voltage (a voltage is detected), we assume that the function has collapsed into a state from which it cannot go back to the zero  \cite{Andersen13}.
In the opposite case, in which the voltage {\it has not} appeared, after the measurement we modify the wavefunction $\psi_1(\varphi,t)$ to obtain the measured wavefunction $\psi_2(\varphi,t)$  located inside the potential well, region $I$ of Fig. \ref{fig:potential}:
\beq
\label{eq:theta}
\psi_2 (\varphi, t) = \sqrt{\frac{1}{1-p\left( t=nT/N  \right)}}\theta \left[ \varphi^* - \varphi  \right] \psi_1 (\varphi,t).
\eeq

\noindent Here the prefactor $\left[ 1-p \left( t=nT/N  \right) \right]^{-1/2}$ ensures the correct normalization .
In the end, at each $n-th$ measurement  at the time $t=nT/N$ we retrieve both a probability of a voltage switch $p\left( t=nT/N  \right)$ given by Eq.(\ref{eq:p}) and an initial wave function given by Eq.(\ref{eq:theta}).
The latter function is inserted into the time dependent Schr\"odinger equation (\ref{eq:JJSEn}) and integrated for a time interval $T/N$, while ramping the current, to reconstruct a discrete 
time history (in the language of quantum measurements \cite{Anastopoulos08}). The purpose of the calculations is to 
determine the PDF of the switching currents, that we name  ${\cal P}_{\gamma_{sw}}\left( \gamma \right)$ \cite{noteP}.

We start noticing that Eq. (\ref{eq:p}) for $n=1$ describes the probability that a switch occurs at the current $\gamma (t)$.
The initial state at $t=0$ ($n=0$) is surely a static, non switched state; in fact  for $\gamma=0$ the periodic boundary conditions of Eq.(\ref{eq:JJSEn}) forbid tunneling.
Thus, for the first time interval, $n=1$ the connection between  the switching current distribution ${\cal P}_{ \gamma_{sw}} \left(  \gamma \right) $ and the probability 
$p \left(t=T/N \right)$ of a switch  reads  $p(t=T/N) = {\cal P} _{\gamma_{sw}}  \left( 1/N \right)$.
For the subsequent measurements, the probability of a switch is conditioned by the probability that a switch has not occurred in all previous measurements, that gives 
the following  rule for $n \ge 2$:
\beq
{\cal P}_{ \gamma_{sw}  }   \left(   n/N \right)  = p(t=nT/N) \prod_{k=1}^{n-1}\left[ 1-p(kT/N)\right].
\label{eq:recursive}
\eeq
The distribution of switching current probability ${\cal P}_{ \gamma_{sw}  }   \left( n/N  \right) $ is the quantity that can be promptly compared with 
experiments \cite{Martinis87,Martinis02,Wallraff03b,Price10,Coskun12,Massarotti15} and it reproduces 
the qualitative features of macroscopic quantum tunnel  in JJ: the appearance of a peak, or a most probable switching current when the potential energy is comparable 
to the quantum fluctuations \cite{Tinkham96}. 
For a detailed comparison with tunnel theory we consider the WKB approximation for the normalized rate $\Gamma$:
\beq
\Gamma (\gamma)  = \frac{\omega_p(\gamma)}{2\pi} \sqrt{120\pi \frac{7.2 \Delta U(\gamma)}{\omega_p(\gamma)}} \times \exp{\left[ -\frac{7.2 \Delta U(\gamma)}{\omega_p(\gamma)}\right] },
\label{eq:WKB}
\eeq
where $\omega_p=  \left( 1-\gamma^2 \right)^{1/4} \ V_0^{1/2} $ is the normalized plasma frequency.
The corresponding probability distribution of the switching currents reads:
\beq
{ \cal P}_{\gamma_{sw}}  \left(\gamma \right)  ={\mathcal N }\,  T \, \Gamma  \left(\gamma \right)  \times \exp{\left[ -T \int_0^{\gamma} \Gamma(x) dx \right]} .
\label{eq:P}
\eeq
(here ${\mathcal N}= 1-\exp\left[-T\int_0^1\Gamma(x)dx \right]$ is the normalizing factor, the details of the derivation are in the Appendix.)
Eq. (\ref{eq:P}) represents the connection between the WKB tunnel rate expression and the effect of the measurements at time intervals $T/N$.

%%%%%%%%%%%%%%%%%%%%%%%%%%%%%%%%%%%%%%%%%%%%%%%%%%%%%%%%%%%%%%%%%%%%%%%%%%%%%%%%%%%%
%%%%%%%%%%%%%%%%%%%%%%%%%%%%%%%%%%%%%%%%%%%%%%%%%%%%%%%%%%%%%%%%%%%%%%%%%%%%%%%%%%%%
%%%%%%%%%%%%%%%%%%%%%%%%%%%%%%%%%%%%%%%%%%%%%%%%%%%%%%%%%%%%%%%%%%%%%%%%%%%%%%%%%%%%%%%%%%%%%%%%%%%
\section{Dynamics of the state after quantum measurements}
\label{Dynamics}
%%%%%%%%%%%%%%%%%%%%%%%%%%%%%%%%%%%%%%%%%%%%%%%%%%%%%%%%%%%%%%%%%%%%%%%%%%%%%%%%%%%%%%%%%%%%%%%%%%%

\begin{figure}
(a)
\centerline{\includegraphics [scale=0.39]{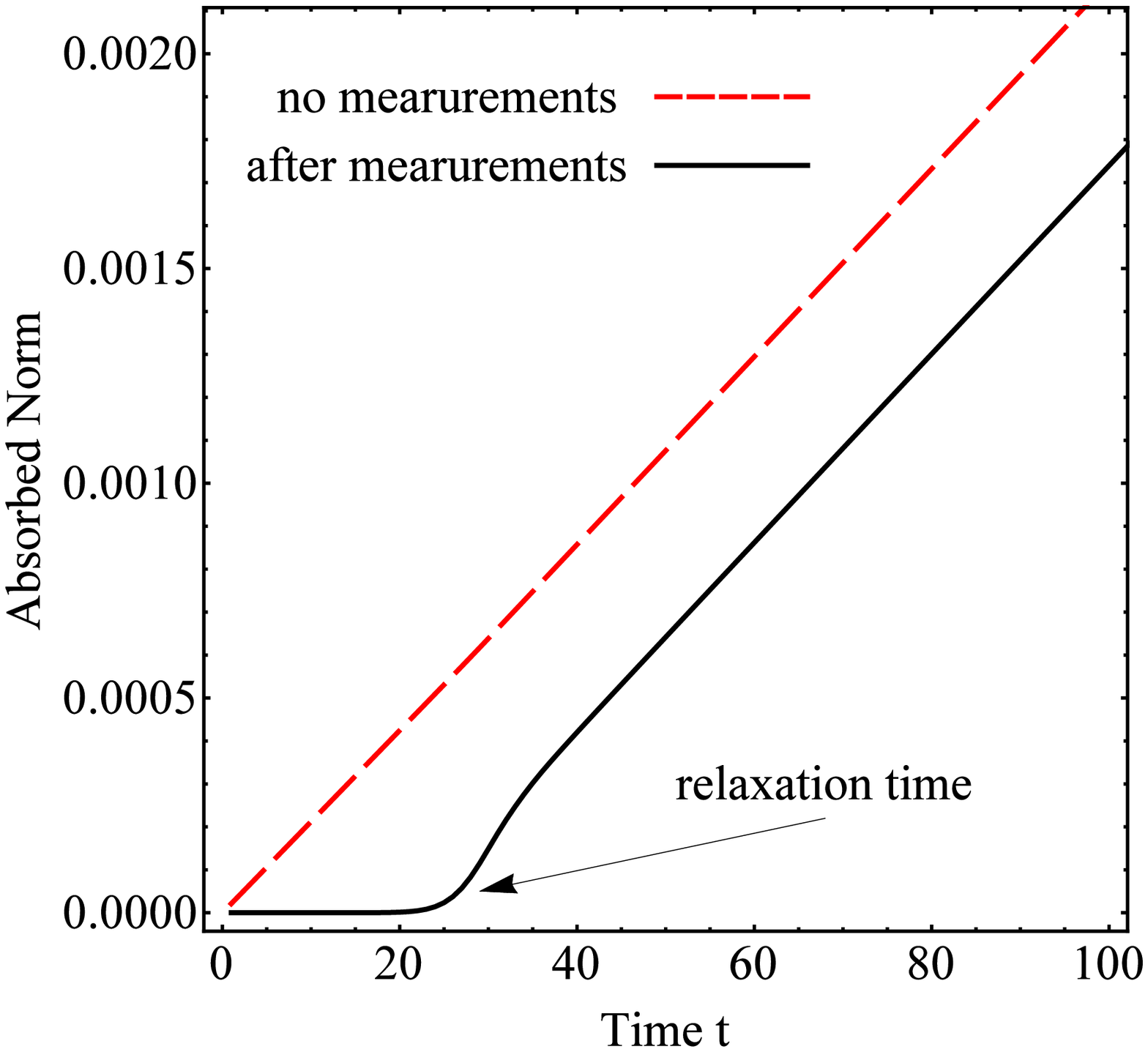}}\\
(b)
\centerline{\includegraphics [scale=0.39]{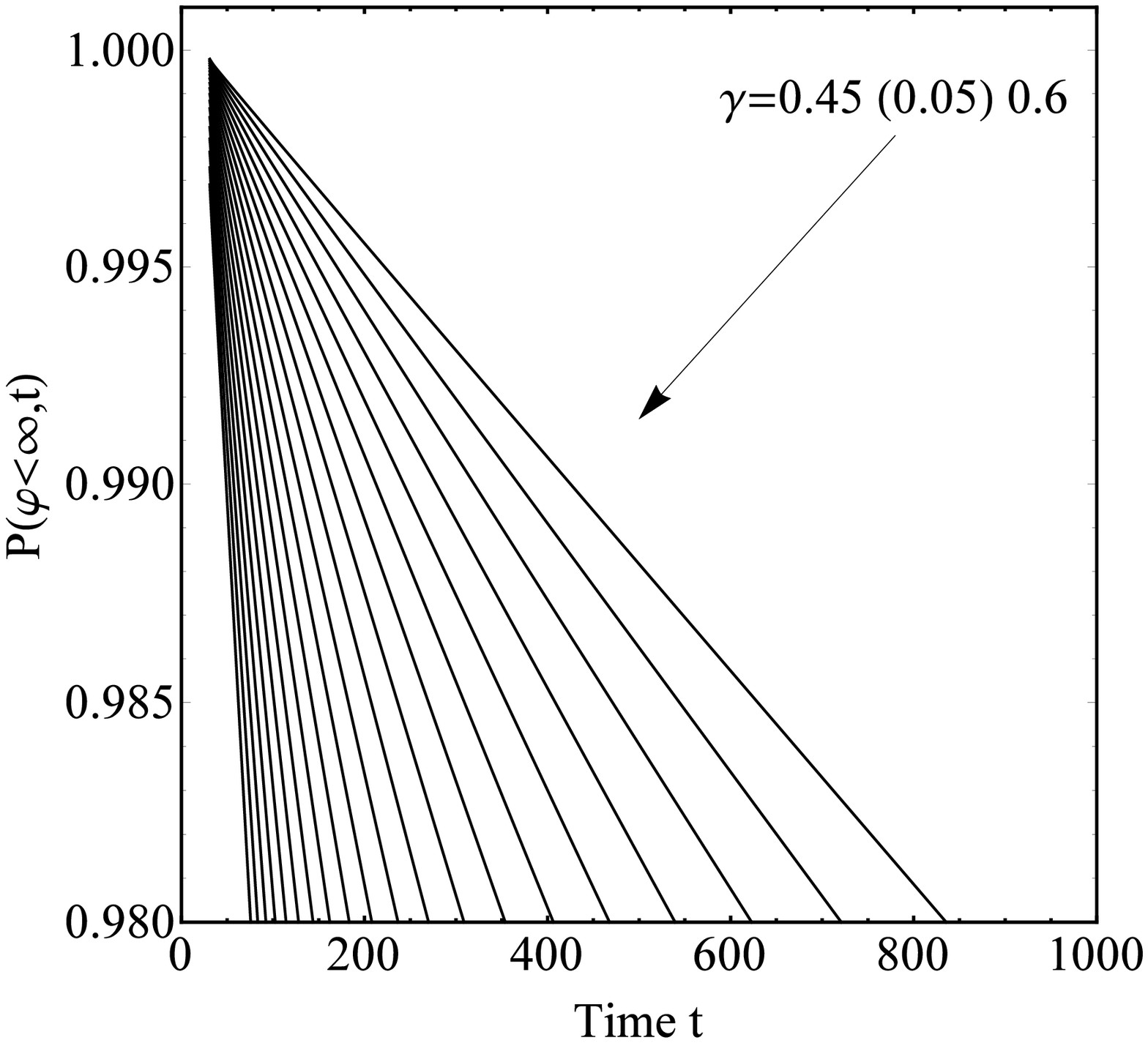}}\\
\caption{(color online). 
(a) Radiated probability absorbed by the radiative boundary condition: red dashed curve, initial condition is the fundamental resonance without measurement; 
black solid line, the  initial conditions is fundamental resonance with measurement as per Eq.(\ref{eq:theta}).\\
(b) 
The probability that the wavefunction is found in the integration domain as a function of the time for different values of the bias current $\gamma$ from $\gamma=0.45$ to $\gamma=0.6$ with step $\gamma=0.05$.\\ 
The normalized maximum energy barrier is $V_0=4$.
}
\label{relaxation}
\end{figure}

In this Section we describe the behavior of the quantum wavefunction after a measurement. 
We only distinguish two outcomes of a voltage measurement in JJ:
\begin{itemize}
\item[{\it i)}]  a running state in the $\varphi > \varphi^*$ region at finite voltage;
\item[ {\it ii)}] a  state localized in the $\varphi < \varphi^*$ region at zero voltage.
\end{itemize}
In the former case {\it i)} the current is registered as a switch, the procedure is completed, and it is not necessary to know the further evolution of the device.
In the latter case {\it ii)} the information obtained by the measurement has modified the wavefunction, as described by Eq.(\ref{eq:theta}).
The dynamics of the absorbed norm just after a measurement is shown in Fig. \ref{relaxation}(a), where the effect at infinity of the radiative boundary conditions is 
displayed in two different cases.
The dashed red line is the behavior of a  purely resonant fundamental state, while the solid black curve refers to the decay of the fundamental resonance state after that a measurement has been performed on.
It can be seen that the change in the wavefunction due to the measurement  first causes the appearance of a finite {\em relaxation} time, about $20$ in Fig. \ref{relaxation}a.
After this time interval a radiative decaying starts sets in, with a slope coefficient close  to the coefficient computed for the non measured fundamental state.

Further  in Fig. \ref{relaxation}(b) we report the asymptotic slope of decaying dynamics of the absorbed norm, that clearly depends upon the bias current.
This is so inasmuch the higher the  bias the lower the energy barrier, and therefore the steeper the decay.
%
%
%%%%%%%%%%%%%____FIG_2____relaxation
%%%%%%%%%%%%%%%%%%%%%%%%%%%%%%%%%%%%%%%%%%
%
%

\begin{figure}
\centerline{\includegraphics [scale=0.39]{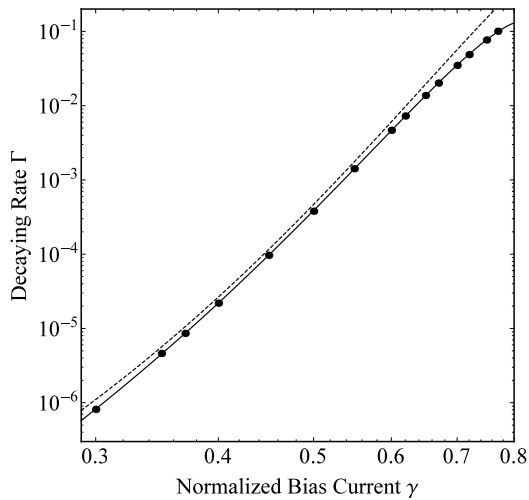}}
\caption{(color online).
Decaying rate of the solution after a quantum measurement (dots-solid line) compared to the WKB rate, Eq.(\ref{eq:WKB}) as a function of the bias current. 
The dimensionless energy barrier reads $V_0=4$.
We underline that both axis are on a log scale. 
%is  energy barrier in Eq.(\ref{eq:WKB}) is 
}
\label{rate}
\end{figure}

\begin{figure}
\centerline{
\includegraphics[scale=0.39]{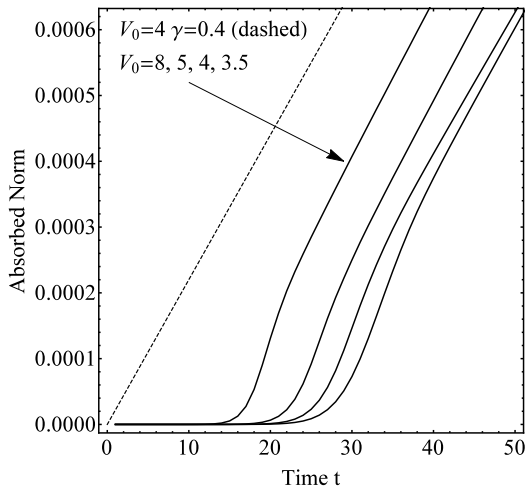}}
\caption{Effect of $V_0$ value  on relaxation time. 
The curves show the behaviour of the absorbed norm at infinity (radiative condition) computed starting from a quantum measured fundamental resonance having the same WKB decaying rate. 
The reference rate is displayed in the dashed line (non measured fundamental state)
for $V_0=4$ and $\gamma=0.4$}
\label{V0_effect}
\end{figure}

In Fig. \ref{rate} we plot the theoretical decaying rate $\Gamma$ as per Eq.(\ref{eq:WKB}) compared to the numerical decaying rates after a quantum measurement, given by the solid line in Fig. \ref{relaxation}(a).
We observe that, apart the constant relaxation time shown in Fig. \ref{relaxation}(a), the observed decaying rate after a measurement well agrees with the theoretical  WKB prediction.
Put it another way, the quantum measurement leaves unchanged the rate and only causes an initial relaxation time. 
This time  depends essentially on the value of barrier height coefficient $V_0$ as elucidated by Fig. \ref{V0_effect}, where different norm absorption dynamics are displayed for the 
same WKB rate (computed for $V_0=4$ and $\gamma=0.4$). The behavior in Fig. \ref{rate} demonstrates that the escape rate weakly  depends upon the bias current far from the 
critical current ($1$ in these normalized units).
Thus, to measure switching currents it is necessary to approach the critical current, since JJ are fast device and the ramp time $T$ is very long in usual measurement setups. 
The quantum measurements effect on switching current should be enhanced for relatively fast ramp time and low $V_0$ barrier coefficient.
This is also reflected in the fact that large changes in the rate cause small changes of the peak of the switching current  distribution.

By way of this part, we have notice that the truncation, see Eq. (\ref{eq:theta}), due to a null measurement induces a delay in the tunnel, for the solution is narrowed in a smaller 
region. Being a constant time lag, the effect becomes negligible when measurements are operated at long intervals, i.e. in the adiabatic regime.

%
%
%%%%%%%%%%%%%____FIG_3____rate_vs_gamma
%%%%%%%%%%%%%%%%%%%%%%%%%%%%%%%%%%%%%%%%%%%%%%%%%%%%%%%%%%%%%%%%%%%%%%%%%%%%%%%%%%%%
%
%
%%%%%%%%%%%%%____FIG_4____V0_effect_relaxation_time
%%%%%%%%%%%%%%%%%%%%%%%%%%%%%%%%%%%%%%%%%%
%

%%%%%%%%%%%%%%%%%%%%%%%%%%%%%%%%%%%%%%%%%%%%%%%%%%%%%%%%%%%%%%%%%%%%%%%%%%%%%%%%%%%%
%%%%%%%%%%%%%%%%%%%%%%%%%%%%%%%%%%%%%%%%%%%%%%%%%%%%%%%%%%%%%%%%%%%%%%%%%%%%%%%%%%%%
%%%%%%%%%%%%%%%%%%%%%%%%%%%%%%%%%%%%%%%%%%%%%%%%%%%%%%%%%%%%%%%%%%%%%%%%%%%%%%%%%%%%
\section{Results}
\label{Results}
%%%%%%%%%%%%%%%%%%%%%%%%%%%%%%%%%%%%%%%%%%%%%%%%%%%%%%%%%%%%%%%%%%%%%%%%%%%%%%%%%%%%
We here collect numerical results and theoretical estimates.
The transient dynamics in between two measurements described in Sect. \ref{Dynamics} suggests that two regimes can be recognized in the quantum escape simulations: a) small time intervals between two measurements, i.e. $T/N$ comparable with the relaxation time; (b) large time lag between measurements, when $T/N$ is large and the relaxation time can be neglected.
We treat the two regimes in the following subsections: In the first case brute force numerical simulations are the only mean to retrieve the escape rate and the current distribution. 
To the contrary, for long time intervals $T/N$ numerical simulations are difficult, inasmuch the integration time becomes prohibitive. 
In this regime we resort to an adiabatic approximation of the dynamics in between two measurements, as will be discussed in Subsect. \ref{ResultsLong}

\subsection{Numerical simulations for small  $T/N$}
\label{ResultsSmall}
%
%%%%%%%%%%%%%%%%%%%%%%%%%%%%%%%%%%%%%%%%%%%
%%%%%%%%%%%%%%____FIG_5___Switch_currents_____
\begin{figure}
\centerline{\includegraphics [scale=0.39]{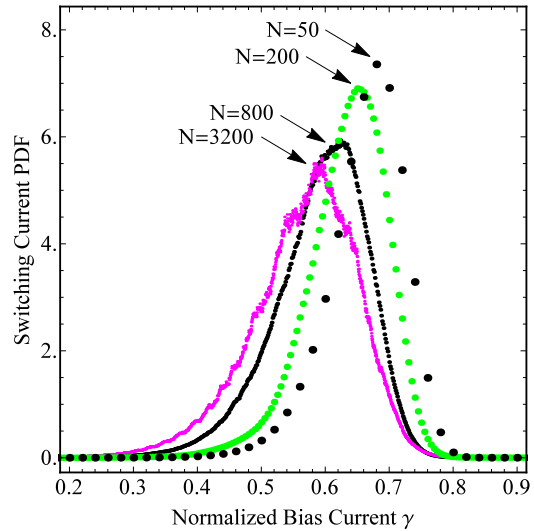}}
\caption{(color online). The probability distribution ${\cal P}_{\gamma_{sw}} \left ( n/N \right)$ of the switching currents  for repeated measurements. The ramp time is $T=800$; during this time the number of measurements changes from $N=50$ to $N=3200$. The normalized maximum energy barrier is $V_0=4$.}
\label{fig:Isw}
\end{figure} 
%%%%%%%%%%%%%%%%%%%%%%%%%%%%%%%%%%%%%%%%%%
%

\begin{figure}
\centerline{\includegraphics [scale=0.39]{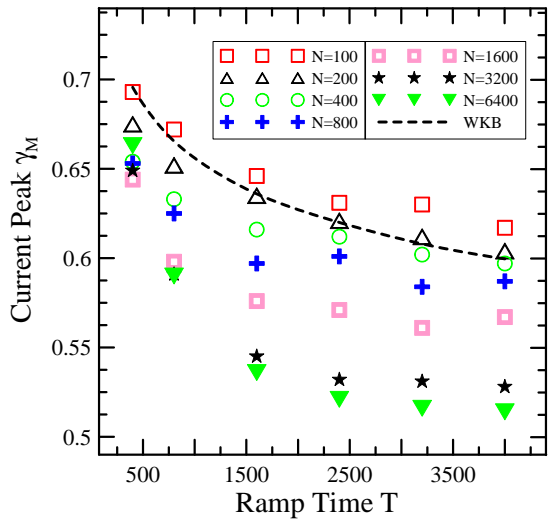}}
\caption{(color online). Behavior of the maximum probability of switch as a function of the current ramp time.  For each ramp-time we report the effect of the number of measurements as per Eq.(\ref{eq:JJSEn}) and (\ref{eq:recursive}). 
The dashed line represents the WKB approximation given by Eq.(\ref{eq:WKB}).
The normalized maximum energy barrier is $V_0=4$.
}
\label{fig:N}
\end{figure}

The usual technique to reveal tunnel in JJ is the PDF of the switching current, that we have retrieved with the methods described in Sect. \ref{switching} and that are displayed in Fig. \ref{fig:Isw}.
Without disturbances, neither classical (intrinsic thermal fluctuations or external noise) nor quantistic (tunnel) the distribution should amount to a $\delta$-function peaked at the critical current $\gamma=1$.
Bell shaped distributions as those displayed in Fig. \ref{fig:Isw}, exhibiting premature switches before the critical current, demonstrate the occurrence of quantum tunnel, if noise and thermal fluctuations are kept at bay.
Several features of the PDF of Fig. \ref{fig:Isw} are worth noticing. 
First, the measurement scheme that we employ reproduces the typical shape of the experiments \cite{Martinis87,Martinis02,Wallraff03,Price10,Coskun12,Massarotti15} and of the WKB theory \cite{Tinkham96}.
Second, it is evident that the number of measurements $N$ in a given ramp time ($T=800$ in this case) has an effect on the PDF, but does not lead to the Zeno paradox \cite{Abdullaev11,Abdo14}.
%Third, the steep $\theta$ function employed to model the nonideal  quantum measurements, see Eq.(\ref{eq:IC}), produces high 
%frequency oscillations in the PDF, that are of course more evident when the number of measurements increases. 
To sum up the effect of the measurements, we focus on the peak of the distribution of the current switches.
The behavior is shown in Fig. \ref{fig:N},  where the current $\gamma_M$ at which a peak of the switching probability occurs is displayed, as a function of the ramp time $T$
 for different values of the number of measurements $N$ (for reference, we also include  the WKB approximation).
From Fig. \ref{fig:N} it is evident that the peak $\gamma_M$ moves to lower currents when the current is ramped more slowly. 
The qualitative  agreement with he WKB approximation (\ref{eq:WKB},\ref{eq:P}) is to be expected, as the slower the ramp the longer it takes the barrier to decrease 
\cite{Fulton74,Kato96, Wallraff03}.
The new element in the calculations of Fig. \ref{fig:N} is the effect of the number $N$ of discrete measurements along the ramp. 
For a given ramp time $T$ the effect of the measurements is to decrease the current at which a peak of the switching probability occurs. 
This is intuitively to be expected, as in the limit of no measurements the system is not observed, and therefore the switching events cannot be registered no 
matter how is high the current. 
It is also interesting to notice that the WKB approximation does not corresponds neither to the limit of infinite measurements ($N\rightarrow \infty$, continuous measurements, 
or monitoring) neither to the case of rare measurements ($N \rightarrow 0$).
We thus confirm that the WKB estimate, Eq.(\ref{eq:theta}), that does not include the effects of the measurements,  is notwithstanding a reliable guess 
of the tunnel induced switching current distribution.

%
%
%%%%%%%%%%%%%____FIG_6____Maximum_proability
%%%%%%%%%%%%%%%%%%%%%%%%%%%%%%%%%%%%%%%%%%
%
%

%%%%%%%%%%%%%%%%%%%%%%%%%%%%%%%%%%%%%%%%%%%%%%%%%%%%%%%%%%%%%%%%%%%%%%%%%%%%%%%%%%%%
\subsection{Numerical simulations for large $T/N$}
\label{ResultsLong}
\begin{figure}
\centerline{\includegraphics [scale=0.39]{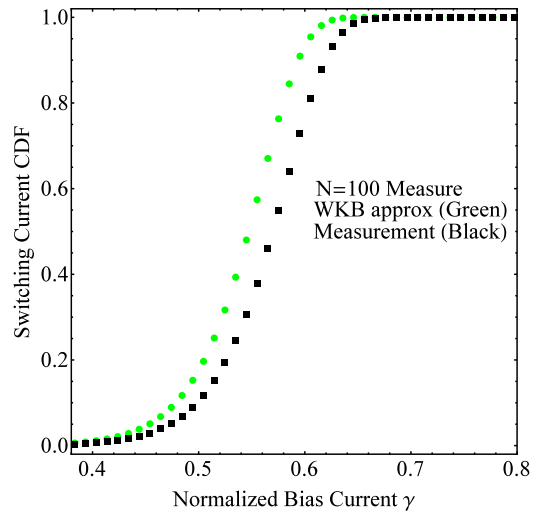}}
\caption{(color online). Cumulative distribution of quantum escape times in the long $T/N$ regime. 
The green dots represent the WKB switching current cumulative distribution corresponding to the PDF reported in Eq.(\ref{eq:P}).
Black dots represent numerical simulations.
The normalized maximum energy barrier is $V_0=4$, the number of measurement reads $N=100$, and the ramp time is  $T=10^4$.
}
\label{fig:asymptoticCDF}
\end{figure}

\begin{figure}
\centerline{\includegraphics [scale=0.39]{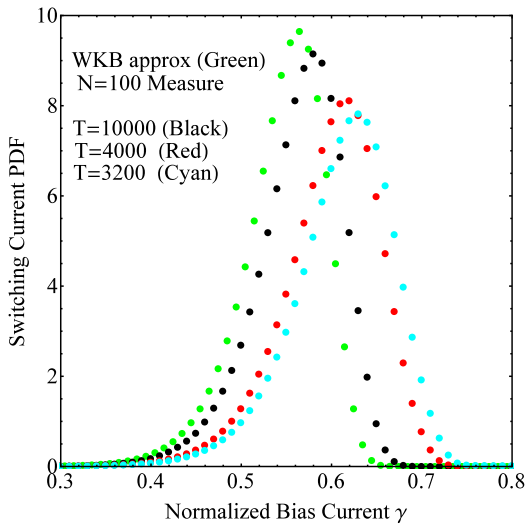}}
\caption{(color online). 
The probability distribution ${\cal P}_{\gamma_{sw}}\left(n/N \right)$ of the switching currents  for repeated measurements and the WKB approximation for 
$T=10^4$. 
The number of measurements is $N=100$ while the ramp time changes from  $T=3200$  to $T=10000$. The normalized maximum energy barrier is $V_0=4$.}
\label{asymptoticConvergence}
\end{figure}

%%%%%%%%%%%%%%%%%%%%%%%%%%%%%%%%%%%%%%%%%%%%%%%%%%%%%%%%%%%%%%%%%%%%%%%%%%%%%%%%%%%%
We here discuss the method to retrieve the switching current distributions for large ratios of the ramp times respect to the number of measurements, namely $T/N$.
% We do so to address the problem of the reliability of the WKB theory, that neglects measurements, in the analysis of actual experiments.
In the limit of long intervals between subsequent measurements, $T/N \rightarrow \infty$,  the relaxation time after each measurement becomes negligible and the 
consequences of wavefunction collapse becomes inessential. Figure \ref{fig:asymptoticCDF} demonstrates that for a time interval $T/N \simeq 10^2$ it is possible to 
appreciate a difference between the WKB approximation and the numerical simulations. The discrepancy between WKB theory (that neglects quantum measurements) and 
experiments (that of course do make measurements) can be ascribed to the perturbation due to the measurements introduced by a finite time lag (the relaxation time, see Fig. \ref{relaxation}).
%%%%%%%%%%%%%____FIG_7____asymptotic
%%%%%%%%%%%%%%%%%%%%%%%%%%%%%%%%%%%%%%%%%%
%
To make a direct comparison with the available experimental data, we have performed simulations of the switching currents to elucidate the convergence towards 
the WKB approximation keeping fixed the number of measurements ($N=100$) and increasing the ramp time in the range $T \simeq 10^3 \div 10^4$, see Fig. \ref{asymptoticConvergence}. 

In this representation is evident that the peak of the distribution moves when the ramp time decreases , approaching the WKB limit.
For relatively slow ramp times, namely $T= 10^4$, there is still a detectable difference with the WKB approximation that neglects the effect of measurements.
The normalized value $10^4$ corresponds, for the typical timescale of the JJ to an accessible ramp time.
Therefore, even if smaller ramp times are not available, the neat difference between the standard approach that neglects measurements 
(the green curves in Figs. \ref{fig:asymptoticCDF} and \ref{asymptoticConvergence}) and the WKB approximation 
(the black curves in Figs. \ref{fig:asymptoticCDF} and \ref{asymptoticConvergence}) makes it realistic to reveal the quantum nature of the measurements.
%%%%%%%%%%%%%%%%%%%%%%%%%%%%%%%%%%%%%%%%%%%
%%%%%%%%%%%%%%____FIG_8___Switch_currents__asymptotic___
%%%%%%%%%%%%%%%%%%%%%%%%%%%%%%%%%%%%%%%%%%
%
%%%%%%%%%%%%%%%%%%%%%%%%%%%%%%%%%%%%%%%%%%%%%%%%%%%%%%%%%%%%%%%%%%%%%%%%%%%%%%%%%%%%
\section{Conclusions}
\label{Conclusions}
%%%%%%%%%%%%%%%%%%%%%%%%%%%%%%%%%%%%%%%%%%%%%%%%%%%%%%%%%%%%%%%%%%%%%%%%%%%%%%%%%%%%

%we have shown that the escape times of a quantum version of the JJ washboard potential are affected by the measurement procedure.
%This influence is a genuine manifestation of the quantum nature of a mesoscopic device and it is independent of the knowledge of the effective noise temperature. addressed the problem to demonstrate the quantum character of a mesoscopic device even if the effective noise temperature is unknown. 
In conclusion, we have modeled the influence of measurements on Josephson mesoscopic devices with a time dependent Schr\"odinger equation, dealing with the problem of the boundary conditions with an appropriated analogue of a perfect matched layer.
Solving this model, we have studied the effect of the  number of measurements and of the bias current ramp time on the switching current distribution.
The effect is evident in the fast ramp time regime, as a consequences of the presence of a relaxation time, but it is also sizable in the slow ramp time regime.
%In fact, in the adiabatic and in the non adiabatic regimes the switching current distributions depend   the probability to observe a voltage switch in correspondence of a certain bias current (the usual signature employed to pinpoint tunnel events) is influenced by the number of measurements. 
In the limit of continuous monitoring our approach reproduces a bell shaped distribution that is compatible with the standard WKB result.
As the number of measurements increases, the peak of the switching current moves towards lower values.
The effect, in the adiabatic regime, is best highlighted when the dimensionless energy barrier $V_0$ is of the same order of magnitude as the lowest energy level, while the shift becomes vanishingly small for larger $V_0$ values.
Thus, the main feature of our findings is that the peak of the switching currents distribution depends on the measurement scheme:
increasing the number of measurements, while keeping the ramp time constant, tunnel events occur at a lower current.
We underline that the change of the switching statistics as a consequence  of measurements is a quantum signature that has no classical counterpart 
and cannot be confused with the effect of undesired noise entering the system.

We feel it is important to mention some limitations of the present work. 
First, we have neglected fluctuations and dissipation, that require a different version of the Schr\"odinger equation \cite{Davidson92,Shnirman97}, as we are chiefly interested in the effect of measurements, not in the details of the dynamics.
Secondly, although we have considered the voltage as the measured quantity, we have not projected the states on the voltage eigenfunctions; we have just used an operational approach of nonideal quantum measurements to qualitatively reproduce the experiments.
Third, the time scale of experiments on superconducting JJ is much longer than numerical simulations, for actual experiments occur on a scale of $\sim 10^9$ normalized units;
To retrieve the effect of realistic ramp time requires a combination of numerical techniques (to obtain the response to the measurement perturbation) and analytical techniques 
(to extrapolate the tunnel rate). 
A natural extension of this work is the use of more refined models, or more extended simulations, to remove the above limits  \cite{Pierro16}.
However, we speculate that the shift of the switching current distribution peak towards lower values, while increasing the number of measurements (as shown in Figs. \ref{fig:N}), could be a robust result.

\section*{Acknowledgements}

We thank A. Davidson, S. Pagano and A. Ustinov for useful suggestions and helpful discussion.
We acknowledge financial support from 
"Programma regionale per lo sviluppo innovativo delle filiere Manifatturiere strategiche della Campania Filiera WISCH, Progetto2: Ricerca di tecnologie innovative digitali per lo sviluppo sistemistico di computer, circuiti elettronici e piattaforme inerziali ad elevate prestazioni ad uso avionico."
VP acknowledges INFN, Sezione di Napoli (Italy) for partial financial support .

%%%%%%%%%%%%%%%%%%%%%%%%%%%%%%%%%%%%%%%%%%%%%%%%%%%%%%%%%%%%%%%%%%%%%%%%%%%%%%%%%%%%
%%%%%%%%%%%%%%%%%%%%%%%%%%%%%%%%%%%%%%%%%%%%%%%%%%%%%%%%%%%%%%%%%%%%%%%%%%%%%%%%%%%%
%%%%%%%%%%%%%%%%%%%%%%%%%%%%%%%%%%%%%%%%%%%%%%%%%%%%%%%%%%%%%%%%%%%%%%%%%%%%%%%%%%%%
\section*{Appendix - Switching Current Distribution in Adiabatic Approximation}
\setcounter{equation}{0}\renewcommand{\theequation}{A\arabic{equation}}

%%%%%%%%%%%%%%%%%%%%%%%%%%%%%%%%%%%%%%%%%%%
%%%%%%%%%%%%%%____FIG_9___prefactor___
\begin{figure}
\centerline{\includegraphics [scale=0.39]{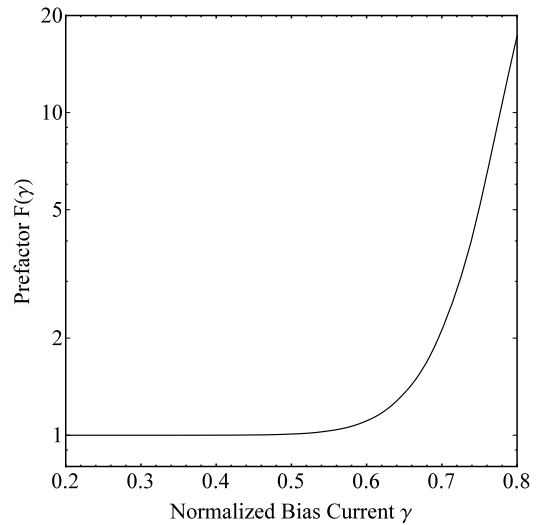}}
\caption{The prefactor relevant for Eq.(\ref{eq:fultonQM})  taking into account 
relaxation time correction to the decaying of measured fundamental resonance. 
The barrier height reads $V_0=4$.}
\label{prefactor}
\end{figure}

The purpose of this Appendix is to apply the adiabatic approximation for the switching current distribution in the limit $T/N \rightarrow \infty$.
We recall that the PDF of the switching currents depends upon the bias sweep rate. In the adiabatic approximation of the quantum evolution of a time dependent 
system the transition occur between instantaneous resonant states.
This can be done, in  our scheme, if the time interval between measurements is greater than the relaxation time. 
The starting point is the theory of Fulton and Dunkleberger \cite{Fulton74} for the tunnel rate from a metastable potential \cite{Kato96,Wallraff03}. 
We start by noticing that a bias threshold $\gamma$ in the range $[0,1]$ is uniquely identified along the ramp by the instant $t$ when $t = \gamma T$.  
If $\Delta \gamma$ is an increment of the bias small enough that the tunnel rate $\Gamma$ of Eq.(\ref{eq:WKB}) can be considered constant, the following formula holds:
\begin{equation}
\mathrm{Prob} \left( \gamma_{sw} > \gamma+\Delta \gamma \right)  = \exp{ \left[ -\Gamma(\gamma) \Delta \gamma T\right]} \, \mathrm{Prob} \left(\gamma_{sw} > \gamma \right).
\label{eq:fulton}
\end{equation}
i.e., the probability to observe a switching current greater  than $\gamma+\Delta \gamma$ is equal to the probability to have jointly a switch above $\gamma_{sw} = \gamma$ and that no escape has occurred in the tiny time interval  $\Delta t = T \Delta \gamma$. 
The exponential factor assumes that the effect of the bias change (during the process the bias rises to $1$ in the time $T$) is negligible in $\Delta t$, and is the probability to have no switching during the small time interval $\Delta t = T \Delta \gamma$ .
Eq. (\ref{eq:fulton}) can be recasted in a differential equation in the limit of infinitesimal $\Delta \gamma$ introducing 
$CDF(\gamma)=1-\mbox{Prob}\left(\gamma_{sw} > \gamma\right )$. 
Furthermore we note that $\mathcal{P}_{\gamma_{sw}}\left(\gamma \right)=\frac{d}{d\gamma}CDF(\gamma)$.
After straightforward manipulations we have:
\begin{equation}
\frac{d}{d\gamma} CDF(\gamma)=\Gamma(\gamma) T (1-CDF(\gamma))
\label{eq:fultoneq}
\end{equation}
with the additional  condition that PDF should be normalized to unity.
The solution of the previous equation, with the specified normalization condition, gives for  $\mathcal{P}_{\gamma_{sw}}(\gamma)$ the following simple expression:
\begin{equation}
\mathcal{P}_{\gamma_{sw}}(\gamma)= \mathcal{N}~ T~\Gamma(\gamma) \times \exp\left [ -T \int_0^\gamma \Gamma(x) dx \right ]
\end{equation}
that is Eq.(\ref{eq:P}) \cite{Fulton74, Kato96, Wallraff03}.
The normalization constant $\mathcal{N}= 1-\exp(-T \int_0^1 \Gamma(x) dx)$ in the adiabatic regime is, roughly, $\mathcal{N} \approx 1$.
 
%The previous formula is Eq.(\ref{eq:P}), and it is used to compute the switching current distribution in the WKB approximation of rate $\Gamma$.

In the quantum measurements regime the discrete histogram giving the distribution of the $N$ measurements in the discrete set of switching current
$\{\gamma_k=\frac{k}{N}\}_{k=1}^{N}$ can be computed starting again from Eq. (\ref{eq:fulton}).
In particular we can write
\begin{eqnarray}
{\rm Prob}(\gamma_{sw} > && \gamma_k)   =  \nonumber \\
 F(\gamma_k) &&  \exp(-\Gamma(\gamma_k) \Delta \gamma T) {\rm Prob} \left(\gamma_{sw} > \gamma_{k-1} \right)
\label{eq:fultonQM}
\end{eqnarray}
where, taking into account the result of intra-measurement dynamic, we have that 
\begin{equation}
P\left( \varphi < \varphi^*,t=\frac{T}{N} \right) = F(\gamma_k) \exp(-\Gamma(\gamma_k) \Delta \gamma T)
\label{eq:noescape}
\end{equation}
that is valid in the limit $T/N$ larger than the relaxation time and in the adiabatic regime.
In Eq. (\ref{eq:noescape}),  $\Gamma(\gamma)$ is the numerical rate and the prefactor $F(\gamma) > 1$, as shown in Fig. \ref{prefactor}, takes into account the effect of  quantum measurements, mainly due to the appearance of a relaxation time.
We use Eq. (\ref{eq:fultonQM}) as a discrete recursive formula to compute 
$\mathcal{P}_{\gamma_{sw}}(\gamma_k)= \mbox{Prob}(\gamma_{sw} > \gamma_{k-1} ) - \mbox{Prob}(\gamma_{sw} > \gamma_{k} )$, with the initial condition $\mbox{Prob}(\gamma_{sw} > \gamma_0 )=1$.

\newpage
%%%%%%%%%%%%%%%%%%%%%%%%%%%%%%%%%%%%%%%%%%%% here start bliblio %%%%%%%%%%%%%%%%%%%%%%%%%%%%%%%%%%%%


\begin{thebibliography}{99}
\bibitem{noteP} For brevity, we use the same symbol ${\cal P}_{\gamma_{sw}}\left( \gamma \right)$ to denote both the continous PDF and the discrete hystogram.
\end{thebibliography}
\end{document}